\documentclass[12pt]{iopart}
\usepackage{graphicx}
\usepackage{iopams}
\usepackage[pdfusetitle,
bookmarks=true,colorlinks=true, linkcolor=blue, citecolor=blue, linktoc=page]{hyperref}
\usepackage{comment}
\usepackage{multicol}
\usepackage{floatpag}
\usepackage{subfig}
\usepackage{booktabs}
\usepackage{amssymb}
\usepackage[numbers]{natbib}
\usepackage{multirow}
\usepackage{amssymb}
\usepackage{pifont}
\usepackage{colortbl}
\usepackage{xcolor}

\begin{document}

\title[]{Tungsten erosion and scrape-off layer transport modelling in L-mode helium plasma discharges in ASDEX Upgrade}

\author{G. Alberti$^{1}$, E. Tonello$^{2,1}$, C. Tuccari$^{1}$, F. Mombelli$^{1}$, S. Brezinsek$^{3}$, T. Dittmar$^{3}$, A. Hakola$^{4}$, A. Kirschner$^{3}$, K. Krieger$^{5}$, M. Rasinski$^{3}$, J. Romazanov$^{3}$, A. Uccello$^{6}$, A. Widdowson$^{7}$, M. Passoni$^{1,6}$, the ASDEX Upgrade Team$^{a}$ and the EUROfusion Tokamak Exploitation Team$^{b}$}

\address{$^{1}$ Politecnico di Milano, Department of Energy, Milan, 20133, Italy}
\address{$^{2}$ École Polytechnique Federale de Lausanne (EPFL), Swiss Plasma Center (SPC), CH-1015 Lausanne, Switzerland}
\address{$^{3}$ Forschungszentrum Jülich GmbH, Institut für Energie- und Klimaforschung—Plasmaphysik, Partner of the Trilateral Euregio Cluster(TEC), Jülich, Germany}
\address{$^{4}$ VTT Technical Research Centre of Finland Ltd, Espoo, Finland}
\address{$^{5}$ Max-Planck-Institut für Plasmaphysik, Boltzmannstr. 2, 85748 Garching, Germany}
\address{$^{6}$ Istituto per la Scienza e Tecnologia dei Plasmi, CNR, Milan, 20125, Italy}
\address{$^{7}$ United Kingdom Atomic Energy Authority, Culham Centre for Fusion Energy, Culham Science Centre, Abingdon, OXON, OX14 3DB, United Kingdom}
\address{$^{a}$ See author list of \textit{U. Stroth et al. 2022 Nucl. Fusion 62 042006}}
\address{$^{b}$ See author list of \textit{E. Joffrin et al 2024 Nucl. Fusion 64 112019}}

\begin{abstract}
Due to its unavoidable presence in thermonuclear DT plasmas and to its peculiar effects on materials, investigating the role of helium (He) in plasma-wall interaction (PWI) in current tokamaks is fundamental. In this work, PWI in L-mode He plasma discharges in ASDEX Upgrade (AUG) is modelled by exploiting simplified analytical approaches and two state-of-the-art codes. SOLPS-ITER is employed both to provide a suitable background plasma for erosion simulations and to interpret diagnostics measurements in terms of He$^{+/2+}$ fraction. In particular, a 50-50\% concentration of the two He ions is found in the proximity of the strike-points, while He$^{2+}$ represents the dominant population farther in the scrape-off layer (SOL). \\
The role of He ion fraction on AUG tungsten divertor erosion is first estimated by means of a simple analytical model and, afterwards, by exploiting ERO2.0, showing the major impact of He$^{2+}$ in common AUG plasma temperatures. ERO2.0 findings are also compared with experimental erosion data in the strike-point region, showing the possible impact of extrinsic impurities on divertor erosion. \\
Finally, the multi-fluid and kinetic approaches employed in this work to simulate W erosion and migration are compared, including W also in SOLPS-ITER modelling. The impact of target boundary conditions on the W source in SOLPS-ITER is investigated, in order to find a good agreement with ERO2.0 estimation. Then, W migration in the two codes is compared, showing a stronger W transport towards the X-point in ERO2.0 compared to present SOLPS-ITER simulations with no drifts. Similar W influx in core could be achieved by reducing anomalous diffusivity in ERO2.0.
\end{abstract}


\section{Introduction}
The understanding and control of plasma-wall interaction (PWI) in present-day tokamaks are crucial aspects to be addressed in view of nuclear fusion exploitation \cite{roth2009}. The high particle and power fluxes expected in future devices could cause severe damage to materials, such as erosion and melting. These processes represent also an impurity source for the plasma, which could increase dilution and cooling and potentially prevent high energy production \cite{smirnov2015}. Finally, eroded impurities can be redeposited in different locations inside the device, forming layers with different properties with respect to the underneath bulk material and increasing fuel retention due to co-deposition \cite{loarer2009}. \\
The presence of helium (He) in the plasma is expected to play a major role in influencing these phenomena \cite{hakola2017}. As a product of D-T reactions, He will be always present as ash in future reactors. Its higher Z with respect to hydrogenic species can enhance material erosion and plasma dilution. Moreover, its peculiar interaction with metallic materials, such as tungsten (W), can cause important morphology changes on plasma-facing components (PFCs) surface, with the formation of ripples, nanobubbles and especially fiber-form nanostructures called fuzz \cite{baldwin2010}. Being able to predict and possibly mitigate these phenomena is thus of paramount importance. \\
Helium plasmas present a peculiar characteristic with respect to hydrogenic ones which is often overlooked, namely the presence of ions with different charge. The He ion fraction is not directly measured by Langmuir probes (LPs) and can be hardly estimated during experiments. However, it is worth noting that ion charge is a fundamental parameter for erosion, since it determines the impact energy of He ions on the surface. Being able to predict this parameter, understanding how it behaves in different plasma conditions, and studying its role in erosion estimates is of great importance for the interpretation of PWI in He plasma. This is also fundamental to estimate the impurity sources into the main plasma, which can severely affect discharge control. Predicting He ion fraction could be important also for D-T plasma discharges, where a 5-10\% He concentration is expected due to fusion reactions. \\
Experimental campaigns in full He plasma have been recently carried out in different tokamaks, such as ASDEX-Upgrade (AUG) \cite{hakola2017} and WEST \cite{tsitrone2022}, mostly aiming at the investigation of PWI in helium. The choice of full He plasma for those experiments is justified by the need to achieve reactor relevant fluences in lower time and to concentrate on the physics caused by He, without the complicating interplay of other plasma species. In particular, pristine and pre-formed W fuzz samples have been exposed in L-mode and H-mode discharges in AUG He plasmas, to study W fuzz formation, growth and erosion \cite{reinhart2022}. W-coated monoblocks have been also exposed to L-mode He plasmas in WEST. Results showed that fuzz creation and erosion strongly depend on local plasma and surface conditions around the outer strike line. Depending on the position, new W fuzz was formed at AUG during helium exposure, or pre-existing W fuzz was completely eroded. At the same time, no W fuzz was formed on the divertor surfaces in WEST \cite{reinhart2022}, highlighting the higher complexity of fuzz dynamics in tokamaks with respect to more idealized studies on linear devices.  \\
In order to face the complexity of the depicted picture and support the interpretation of experimental findings, numerical models represent a suitable tool. In the case of He plasmas, estimating the He ion fraction and its impact on PFC erosion is almost only possible with modelling. Moreover, testing available tools in different plasma species with respect to hydrogen isotopes and developing complementary and simplified analytical descriptions are of interest to strengthen their interpretative and predictive capabilities for the design of future tokamaks, such as ITER and DTT \cite{romanelli2024}. \\
In this work, the modelling of edge plasma and PWI is presented for an attached L-mode He plasma discharge in the AUG tokamak, with experimental conditions specifically tailored in the way that clear erosion would be obtained for experimental quantification. Two state-of-the-art codes are used for this purpose: SOLPS-ITER \cite{wiesen2015, bonnin2016}, a 2D fluid-plasma and kinetic-neutral solver which assumes diffusive anomalous cross-field transport, and ERO2.0 \cite{romazanov2017}, a 3D Monte-Carlo and impurity transport code. The former is used to properly interpret experimental findings in terms of He$^{+/2+}$ ion fraction and to provide a suitable plasma background for ERO2.0. The validation of these results against experimental data in AUG is presented in section \ref{sec:SOLPSvalidation}. The influence of He$^{+/2+}$ fraction on W erosion is first estimated using a simple analytical model, described in section \ref{sec:analyticalModel}. Then, SOLPS-ITER ion fractions, together with density, velocity and temperature distributions are provided to ERO2.0 to study AUG divertor erosion and the subsequent migration of impurities, which are discussed in section \ref{sec:EROresults}. The obtained results are compared to the two simplified cases of full He$^+$ and full He$^{2+}$ plasma, to highlight the role of the correct ion fraction in erosion evaluation. The impact of parameters which are often unknown in tokamak discharges, such as oxygen presence in plasma and uncertainties in LP temperature measurements, is addressed in comparison to experimental data. Section \ref{sec:Wtransport} discusses the inclusion of W impurities coming from the divertor region in SOLPS-ITER modelling. The impact of boundary conditions on the estimation of W sources in SOLPS-ITER simulations is investigated, and the outcomes are compared to the corresponding predictions of the analytical model and ERO2.0. The SOLPS-ITER multi-fluid approach and the ERO2.0 kinetic one are compared in terms of W transport as a function of diffusivity. Finally, conclusions are drawn in section \ref{sec:conclusions}. 

\section{Validation of SOLPS-ITER plasma background against AUG experiments}
\label{sec:SOLPSvalidation}
\begin{figure}
\centering
\includegraphics[width=0.8\textwidth]{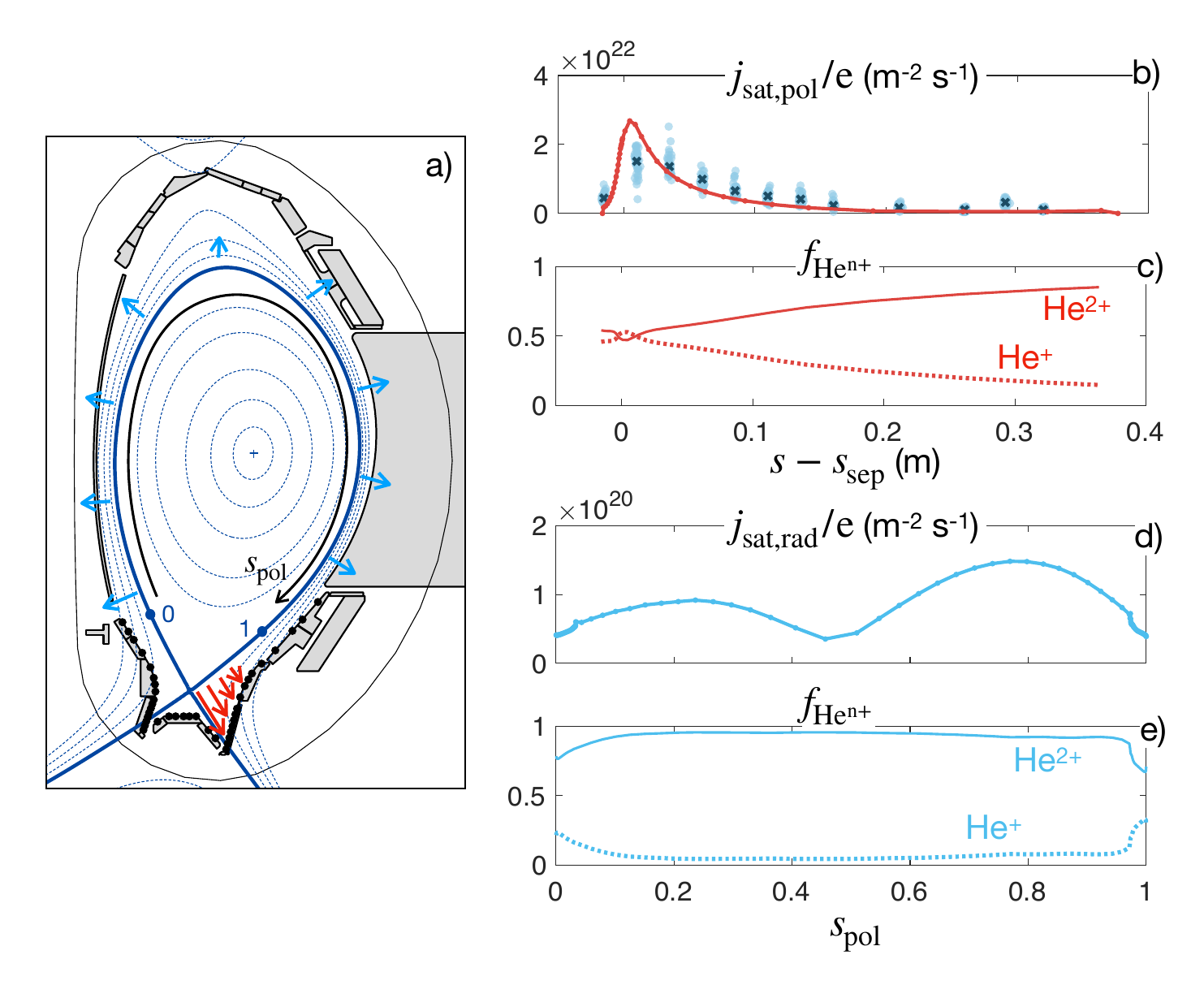}
\caption{(a) Total ion flux on the target (red) and in the far SOL (blue), represented as ion saturation current. (b) Poloidal ion saturation current profiles on the outer divertor target and (c) the corresponding ion flux composition. (d) Radial far SOL ion saturation current and (e) the corresponding ion flux composition.}
\label{fig:SOLPSfluxes}
\end{figure}

This section presents the validation of SOLPS-ITER simulations against experimental data from the L-mode reference AUG discharge $\#36687$. In this experiment, a reversed toroidal magnetic field (namely with $\nabla B$ drift pointing upwards) of $- 2.5$ T is adopted, with a plasma current of $790$ kA. Helium is puffed from the outboard midplane (OMP). Plasma is heated by Ohmic and electron cyclotron resonance heating (ECRH), with power levels of $0.307$ and $1.323$ MW, respectively. The comparison focuses on the OMP and the outer divertor region and the selected diagnostics for comparison are the AUG Integrated Data Analysis (IDA) system \cite{fischer2010integrated}, primarily based on Thomson scattering, and divertor Langmuir probes, respectively. \\
These comparisons are crucial for ensuring the accuracy of the SOLPS-ITER plasma background, which is subsequently used for erosion and impurity transport modelling.
\begin{figure}
\centering
\includegraphics[width=1.0\textwidth]{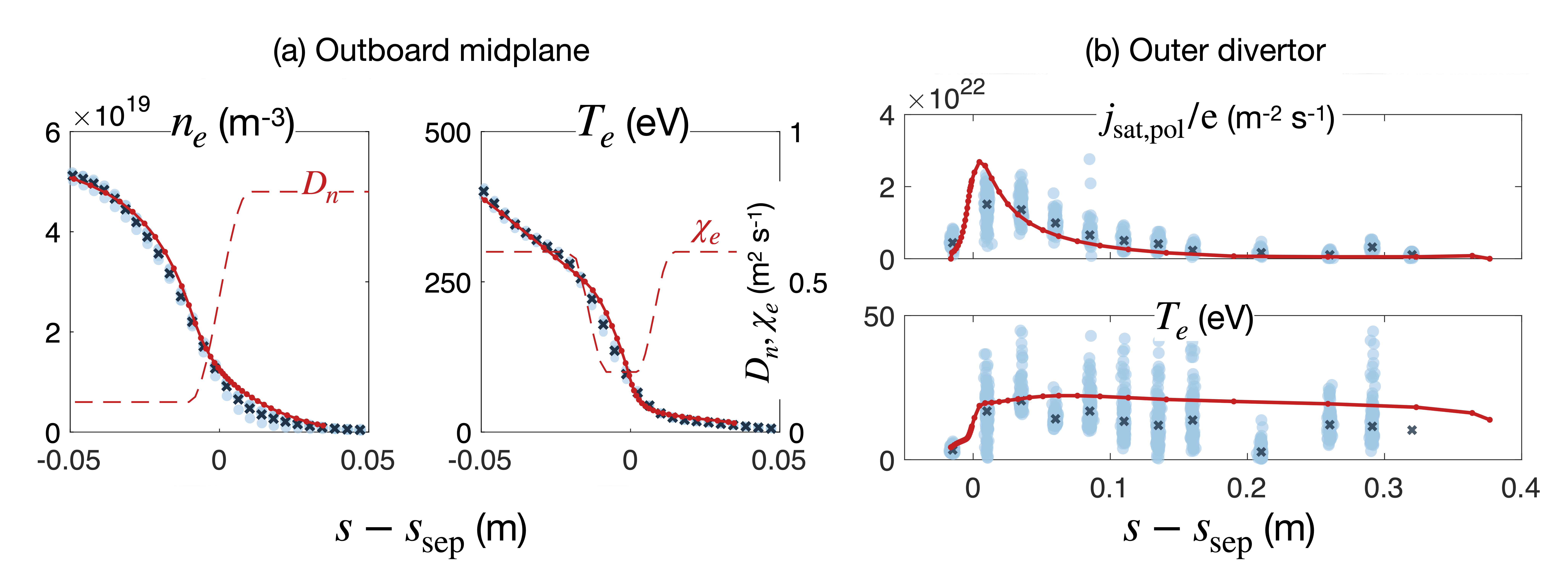}
\caption{Results of the optimised SOLPS-ITER simulations (red line) compared to AUG experimental data (scatter blue dots and the corresponding median (x)). (a) Outboard midplane profiles of electron density and temperature (solid) and optimised profiles of the anomalous transport coefficients (dashed) with respect to experimental data. (b) Outer divertor (normalised) ion saturation current and electron temperature profiles.}
\label{fig:SOLPSexp}
\end{figure} \\
Details on the plasma background simulation setup are given in~\ref{app:SOLPSsetup}. In this section we discuss the results obtained with the optimised set of input parameters, in terms of profile and values of the anomalous diffusivities of cross-field transport. We refer to Chapter 6 of~\cite{TonelloPhD2023} for all the preliminary parametric scans which led to this optimisation. \\
Figure~\ref{fig:SOLPSexp}.a shows the outboard midplane electron density and temperature profiles from SOLPS-ITER simulation (solid red line) compared to the IDA data, acquired between 4.1 s and 4.3 s during the plasma flattop (dots) and the corresponding median values within this interval (crosses). The good agreement is obtained by setting the value of the electron density at the OMP separatrix $n_{e,\mathrm{sep@OMP}} = 1.25 \times 10^{19} \, \mathrm{m^{-3}}$, through density feedback, and then shaping the profile of the anomalous diffusivities $D_n$, for density, and $\chi_e$ and $\chi_i$, for electron and ion heat transport. The profiles of the optimised coefficients to obtain the best agreement with experiments are shown with a dashed line in figure~\ref{fig:SOLPSexp}.a.\\
The simulation results at the outer divertor are plotted in figure~\ref{fig:SOLPSexp}.b. Here, the two quantities under analysis are the ion saturation current, normalised to the unit charge $\mathrm{e}$, $j_\mathrm{sat,pol}/\mathrm{e}$, and the electron temperature. Simulations (solid red line) are compared to the LP measurements taken at the same time interval considered for IDA data. Blue crosses show the median of the experimental data. \\
In helium plasma discharges, estimating the particle flux from the ion saturation current measured by the divertor LPs is not straightforward. Indeed, each $\mathrm{He}^{2+}$ ion collected contributes twice to the electrical current density as each $\mathrm{He}^+$ and knowing the plasma composition is thus crucial. To validate simulations against experiments, we compared the experimental ion saturation current $j_\mathrm{sat}^\mathrm{exp}/\mathrm{e} = I/ (\mathrm{e} \times A_\mathrm{probe}$) to the one estimated by SOLPS-ITER as $j_\mathrm{sat}^\mathrm{SOLPS}/\mathrm{e} = \Gamma_\mathrm{He^+} + 2\times \Gamma_\mathrm{He^{2+}}$, where $\Gamma_\mathrm{He^+}$ and $\Gamma_\mathrm{He^{2+}}$ are the particle fluxes of the two species. Conversely, if from experiments we can only retrieve the total charged flux hitting the divertor targets, multi-ion SOLPS-ITER modelling allows to decouple the key information on the ion flux composition. This is addressed in figure~\ref{fig:SOLPSfluxes}, where the poloidal fluxes on the outer divertor (top) and the radial fluxes towards the first wall (bottom) are analysed. We can first notice that the former and the latter differ by approximately two orders of magnitude. However, the cross-field fluxes are also influenced by the choice of the anomalous particle diffusivity. A direct experimental measure of the magnitude of the radial ion loss, e.g. through LP measurements on the inner or outer limiters, would allow to better constrain this model in terms of absolute magnitude of the cross-field fluxes, for which a predictive theoretical model is not yet available. \\
A remarkable finding of this analysis is that most of the overall particle flux, i.e. the one that impacts the divertor in the region of the strike point, is an approximately equal mixture of $\mathrm{He^{2+}}$ and $\mathrm{He^{+}}$ ions. The high fraction of $\mathrm{He^{+}}$ is due to the recycled neutral atoms that are ionized in the divertor leg in front of the target plates. The radial particle flux, instead, is almost entirely carried by $\mathrm{He^{2+}}$ ions coming from the hot and fully-ionised core region. The impact of the ion flux composition on outer divertor erosion is addressed in the following sections. 

\section{A simple analytical model to estimate the impact of He ion fraction on W erosion}
\label{sec:analyticalModel}
\begin{figure}
\centering
\includegraphics[width=1\textwidth]{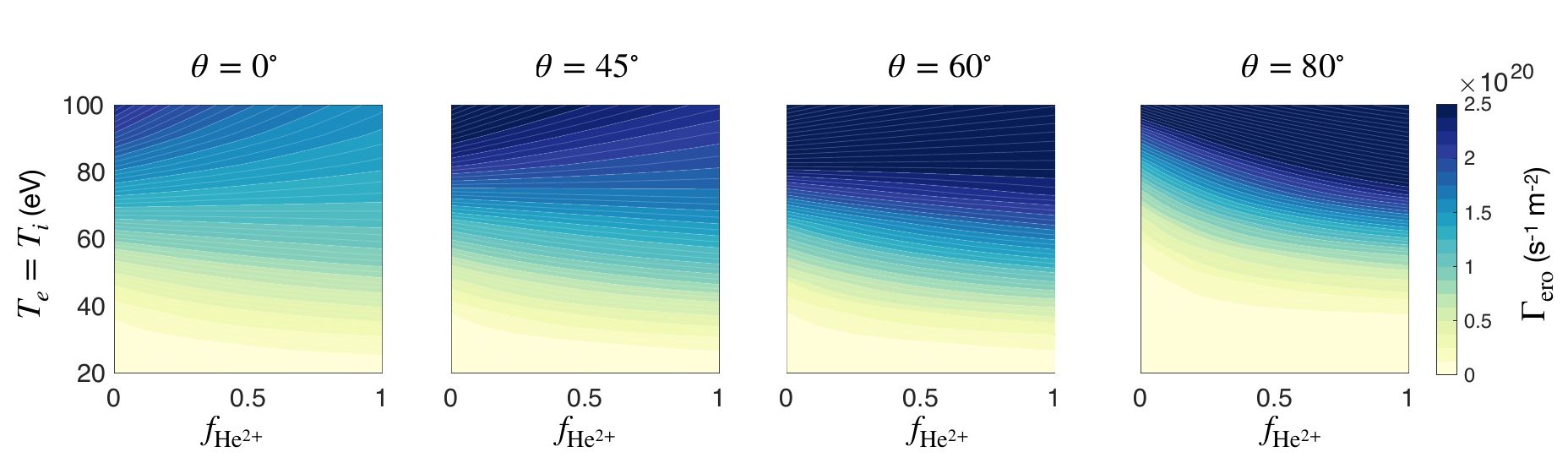}
\caption{Variation of the erosion flux as function of the He$^{2+}$ fraction in plasma $f_\mathrm{He^{2+}}$ and of the plasma temperature $T_e = T_i$ for different incidence angles $\theta$. The saturation current $j_{\mathrm{sat}}/e$ is equal to $1.0 \times 10^{22}$ m$^{-2}$ s$^{-1}$ in all cases, which is comparable to SOLPS-ITER predicted one in figure \ref{fig:SOLPSfluxes}. Level curves for the erosion flux are shown for a better readability of colour variation.}
\label{fig:He_analyticalModel}
\end{figure}

In a helium plasma, the nature of the flux composition, i.e. the fraction of $\mathrm{He}^{+}$ and $\mathrm{He}^{2+}$ ions, greatly affects the erosion of PFCs. The aim of this section is thus to provide an analytical tool based on simplified hypotheses to investigate the role of the two He ions on W erosion. \\
The fraction $f_{\mathrm{He}^{2+}}$ of He$^{2+}$ ions in the plasma can be defined as:
\begin{equation}
f_{\mathrm{He}^{2+}} = \frac{\mathrm{He}^{2+}}{\mathrm{He}^{+} + \mathrm{He}^{2+}} 
\end{equation}
where He$^{+}$ and He$^{2+}$ denote the densities of each ion species. Thus, $f_{\mathrm{He}^{2+}}$ ranges between 0 (full He$^{+}$ plasma) and 1 (full He$^{2+}$ plasma). As mentioned in section \ref{sec:SOLPSvalidation}, a quantity that is usually possible to measure by Langmuir probes in experiments is the saturation current density $j_{\mathrm{sat}}$ at the target, which ultimately represents the total charge reaching the target per unit time and unit area. From this measure, it is not possible to directly discriminate the nature of the flux in terms of He ions. In the two extreme cases, the particle flux on the target will be equal to $j_{\mathrm{sat}}$ for a full He$^{+}$ plasma and equal to $0.5 j_{\mathrm{sat}}$ for full He$^{2+}$, since, in the latter case, each ion is transporting a double charge. \\
The energy of ions impinging on the wall, $E_{\mathrm{wall}}$, is calculated as:
\begin{equation}
E_{\mathrm{wall}} = 2 T_i + Z \mathrm{e}|V_{\mathrm{sh}}| = 2 T_i +  \frac{Z\mathrm{e}}{2} \left| \ln{\left[2 \pi \frac{m_e}{A m_p}\left( 1 + \frac{T_i}{T_e}\right)\right]} \right| T_e
\end{equation}
where $|V_{\mathrm{sh}}|$ represents the sheath potential drop, often approximated as $3T_e$, and $Z \mathrm{e}$ is the ion charge \cite{stangeby2000plasma}. To simplify the analysis further, we will consider $T_i = T_e$. The model, however, could easily be adjusted to account for different temperatures. Considering Eckstein sputtering yields $Y_{\mathrm{He-W}} (E, \theta )$ \cite{behrisch2007sputtering} and assuming the same incidence angle $\theta$ for the two ions, the overall flux of eroded particles due to physical sputtering can be written as:
\begin{equation}
\Gamma_{\mathrm{ero}} = \left [ \frac{f_{\mathrm{He}^{2+}}}{1+f_{\mathrm{He}^{2+}}} Y_{\mathrm{He-W}} (E_{\mathrm{wall, He}^{2+}}, \theta ) + \frac{1-f_{\mathrm{He}^{2+}}}{1+f_{\mathrm{He}^{2+}}} Y_{\mathrm{He-W}} (E_{\mathrm{wall, He}^{+}}, \theta )  \right ] j_{\mathrm{sat}}/e
\end{equation}
which results, keeping $j_{\mathrm{sat}}$ fixed, in $1/2 j_{\mathrm{sat}} Y_{\mathrm{He-W}} (E_{\mathrm{wall, He}^{2+}}, \theta )$ for a full He$^{2+}$ plasma ($f_{\mathrm{He}^{2+}} = 1$), and $ j_{\mathrm{sat}} Y_{\mathrm{He-W}} (E_{\mathrm{wall, He}^{+}}, \theta )$ for a full He$^+$ one ($f_{\mathrm{He}^{2+}} = 0$). \\
A 2D map of the eroded particles flux $\Gamma_\mathrm{ero}$ as a function of $f_{\mathrm{He}^{2+}}$ and $T_e$ is displayed in figure \ref{fig:He_analyticalModel}, for different values of the incident angle $\theta$. As expected, for each $f_{\mathrm{He}^{2+}}$, $\Gamma_\mathrm{ero}$ increases as temperature grows. However, focusing on the graph for normal incidence $\theta=0^{\circ}$, two different regimes can be observed according to plasma temperature, which result in an opposite behaviour of the erosion flux as the fraction of He$^{2+}$ ions increases. For $T \lesssim 65$ eV, the erosion flux increases with $f_{\mathrm{He}^{2+}}$, while the contrary occurs for  $T > 65$ eV. This trend can be ascribed to the competition between incident ion flux and sputtering yield variation. Indeed, at low $T$, sputtering yield presents a steep increase with energy, which dominates incident flux reduction as $f_{\mathrm{He}^{2+}}$ increases. For high $T$, instead, sputtering yield reaches a plateau and could even slightly decrease, leading to similar yields for the two He ions and, ultimately, to an erosion reduction together with the incident ion flux as $f_{\mathrm{He}^{2+}}$ increases. 
 \\
It should be noted that the $65$ eV threshold is specific of the He-W combination at orthogonal incidence, and it could vary according to the energy and angle distributions of the sputtering yield curve for other projectile-target combinations. Moreover, the assumption of same incidence angle for the two He ions could be questioned, since a higher deflection in the sheath could be expected for ions with higher charge. This trend, anyway, appears to be quite general for cases similar to this, namely with a fixed measured $j_{\mathrm{sat}}$. 

\section{ERO2.0 investigation of AUG divertor erosion in He plasma}
\label{sec:EROresults}
This section focuses on the estimation of AUG divertor erosion during He plasma discharges exploiting ERO2.0 capabilities. After a description of the simulation setup, which is necessary to understand following results, the impact of the He ion fraction on AUG divertor erosion is evaluated with ERO2.0. The section then concludes comparing obtained results with experimental findings.

\subsection{ERO2.0 simulation setup}
\begin{figure}
\centering
\includegraphics[width=1.0\textwidth]{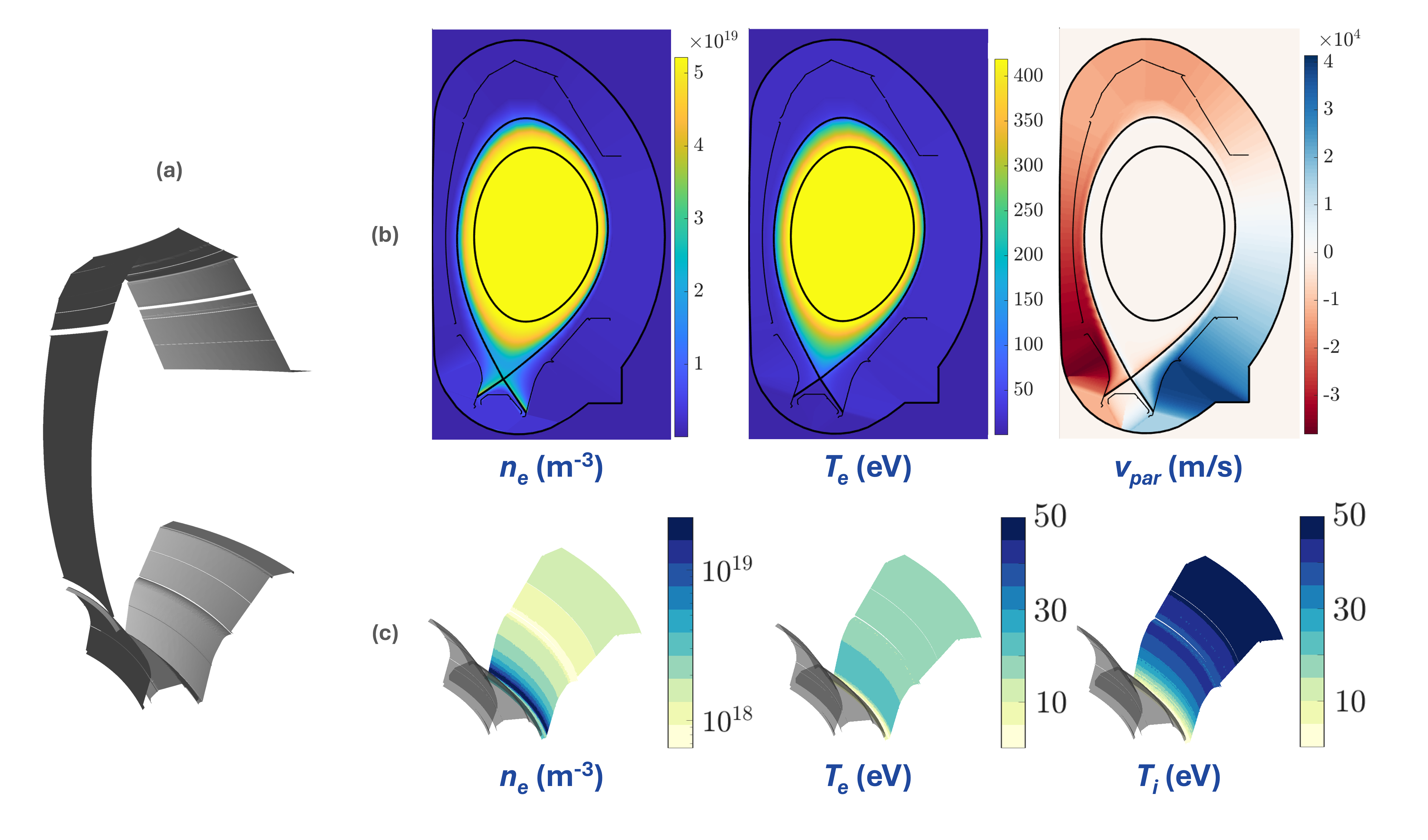}
\caption{Main inputs of ERO2.0 simulations: (a) 30° toroidally symmetric wall, drawn in CAD extruding the 2D SOLPS-ITER poloidal section of the PFCs; (b) 2D poloidal profiles of plasma parameters relevant for impurity migration as imported in ERO2.0; (c) Plasma parameters on outer divertor surface as obtained from extrapolation of SOLPS-ITER results.}
\label{fig:EROinput}
\end{figure}

The AUG 3D geometry reported in figure \ref{fig:EROinput}(a) is a 30° toroidal sector drawn with a CAD tool extruding the 2D SOLPS-ITER poloidal section of the PFCs reported in figure \ref{fig:SOLPSfluxes} and assuming toroidal symmetry. Plasma parameters, in terms of electron density and temperature, ion temperature and parallel velocity, are retrieved from SOLPS-ITER and imported into the ERO2.0 grid. With the exception of the strike-point region, where the SOLPS-ITER computational domain is already in contact with the solid wall, a constant extrapolation has been used to extend SOLPS-ITER plasma parameters up to the PFC surfaces. In the divertor region, this assumption mostly impacts the upper part of outer target, where the SOLPS-ITER grid is few mm up to several cm away from the surface, and yields an upper limit to W erosion there. However, especially in L-mode discharges, this region is not significant for sputtering studies. The main plasma parameters extrapolated on the outer divertor surface are reported in figure \ref{fig:EROinput}(c). Since ERO2.0 is not able to simultaneously acquire more than one ion species density in the volume, effective charge has been used in the plasma volume to account for the presence of the two He ion species, defined as:
\begin{equation}
Z_{eff} = \frac{\sum_i n_i Z_i^2}{n_e} .
\end{equation}
This means that, for each cell of the plasma volume, ERO2.0 assumes a single ion species with charge equal to $Z_{eff}$ and distributed as the electron density. For the plasma particles impacting on PFCs, instead, the fluxes of the two He ions are directly imported from SOLPS-ITER results. The ion incidence energy is sampled from a Maxwellian distribution around the average value of E$_{\mathrm{mean}} = 2 T_i + 3 q T_e$, with $q$ being the He ion charge, while the incidence angle is varied as a free parameter, since no information about it is available from SOLPS-ITER modelling. Sputtering yields for He on W are obtained from previous SDTrimSP \cite{mutzke2024sdtrimsp} simulations, already available in ERO2.0 database \cite{romazanov2024validation}. No other impurities are considered in the plasma, unless otherwise specified. Eroded W is traced in the plasma using the test particle approximation, namely particles are not influencing neither each other nor the background plasma, and adopting the Adaptive Guiding Center Approximation (AGCA) reported in \cite{rode2022}. Ionization and recombination are modelled using the ADAS rate coefficients. Unless otherwise specified, W particles are launched only from the main area of interest to improve statistics, namely the outer divertor region, where experimentally plasma-wall interactions are the strongest. This means that the influence of impurities eroded from other components, like the first wall, on erosion at the outer divertor is neglected in this study. Although assessing the impact of first wall impurities on divertor erosion would require a dedicated modelling campaign, a rough estimate using constant extrapolation of SOLPS-ITER plasma parameters on surface led to a gross erosion of about $5\%$ higher than simulations neglecting main chamber erosion. Extracting charge-exchange neutral fluxes from SOLPS-ITER and investigating a proper extrapolation of the far SOL plasma parameters are necessary aspects to be addressed for a better estimation of first wall erosion \cite{romazanov2021sensitivity}. Moreover, only a small fraction of W test particles is transported from the first wall to the strike-point region, leading to the necessity of finding ways to improve statistics \cite{rode2024multi}. Thus, a more detailed study of first wall erosion and its impact on the divertor exceeds the scope of this work. Eroded W is traced also from the inner divertor region only in simulations shown in section \ref{subsec:SOLPS-ERO_Wdensity}, for a better comparison between SOLPS-ITER and ERO2.0 predictions of W density in plasma. 
 \\

\subsection{Impact of He ion fraction on AUG divertor erosion}
\begin{figure}
\centering
\includegraphics[width=1.0\textwidth]{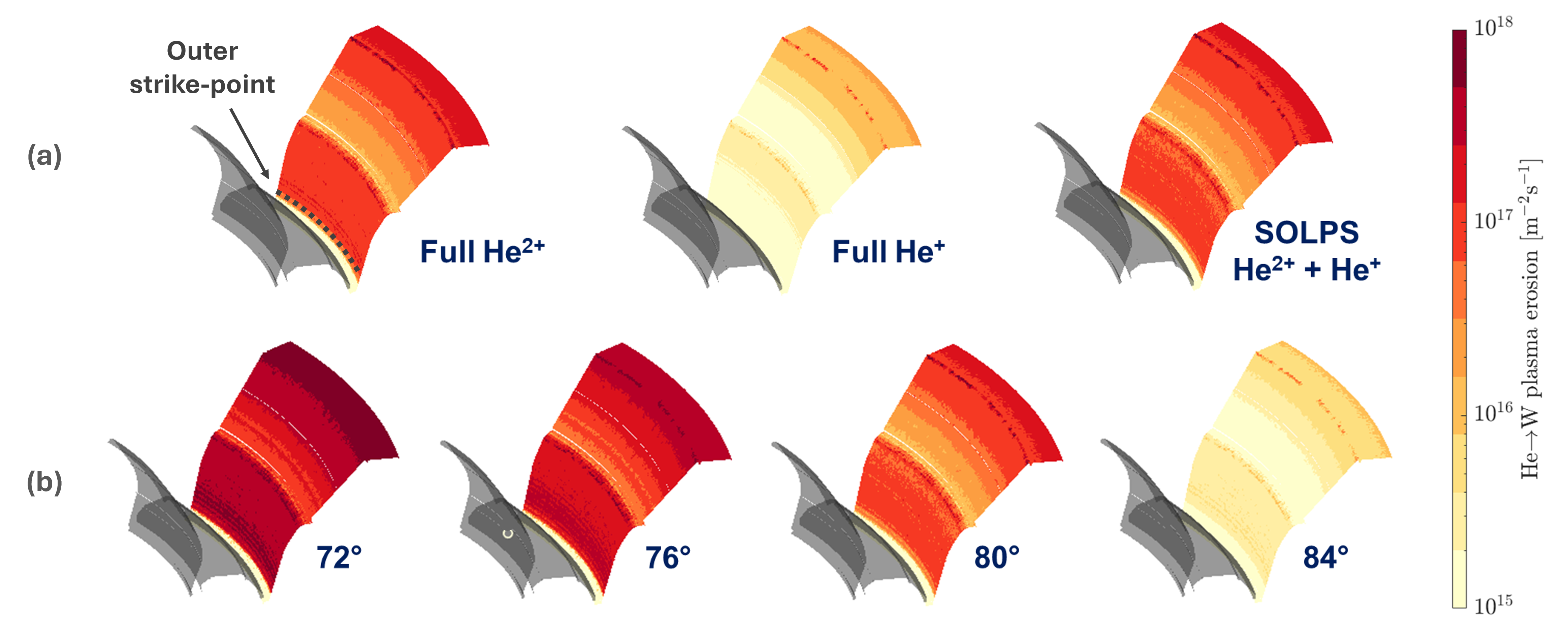}
\caption{AUG outer divertor gross erosion in L-mode He plasma as function of (a) ion fraction (assuming an impact angle of 80° for both He$^+$ and He$^{2+}$) and (b) ion impact angle, as computed by means of ERO2.0 modelling. }
\label{fig:AUG_HePlasmaResults}
\end{figure}

To investigate the impact of He ion fraction on divertor erosion, three different ERO2.0 simulations have been performed, assuming an incidence angle of $80^\circ$. This value has been estimated exploiting the sheath tracing module available in ERO2.0, simulating He$^{2+}$ ions entering the sheath region with $89^\circ$ incidence, which represents the median magnetic field incidence angle on divertor surface in these discharges. With these assumptions, a distribution of incidence angles is found, peaked at about 80°. The same incidence angle is assumed for both He ions, which could represent a slight overestimation of He$^+$ erosion contribution due to its expected lower deflection with respect to He$^{2+}$. The most accurate case, namely with He ion fraction from SOLPS-ITER simulations, has been compared to the two extreme assumptions of full He$^{2+}$ and full He$^+$ plasma, keeping the overall electron density $n_e$ fixed. This allows to directly compare the most accurate description obtained with SOLPS-ITER modelling to the two extreme conditions that could be assumed starting from the same LP experimental data. \\
The results of this analysis, in terms of outer divertor gross erosion, are shown in figure \ref{fig:AUG_HePlasmaResults}(a). The gross erosion found in the most accurate case with SOLPS-ITER ion fractions is comparable with the one obtained with full He$^{2+}$ assumption. In particular, the strongest difference is found in the proximity of the strike-point, where the full He$^{2+}$ case results in a $30\%$ higher W erosion, gradually decreasing moving farther in the SOL. This is consistent with the He$^+$ density obtained with SOLPS-ITER, which concentrates near the strike-point (see figure \ref{fig:SOLPSfluxes}(c)) and contributes more to sputtering yield reduction in that region. On the contrary, full He$^+$ assumption leads to more than one order of magnitude lower gross erosion with respect to the most accurate case.  
 \\
The observed erosion behaviour is in agreement with the outcomes of the analytical model described above. In the AUG conditions investigated here, indeed, both $T_e$ and $T_i$ are lower than the 65 eV threshold for all surface cells. This means that erosion increases as $f_{\mathrm{He}^{2+}}$ increases and it is overall dominated by He$^{2+}$ contribution, even in regions where He$^+$ density approaches $50\%$. \\

\subsection{Comparison between ERO2.0 and experimental findings}
\begin{figure}
\centering
\includegraphics[width=1.0\textwidth]{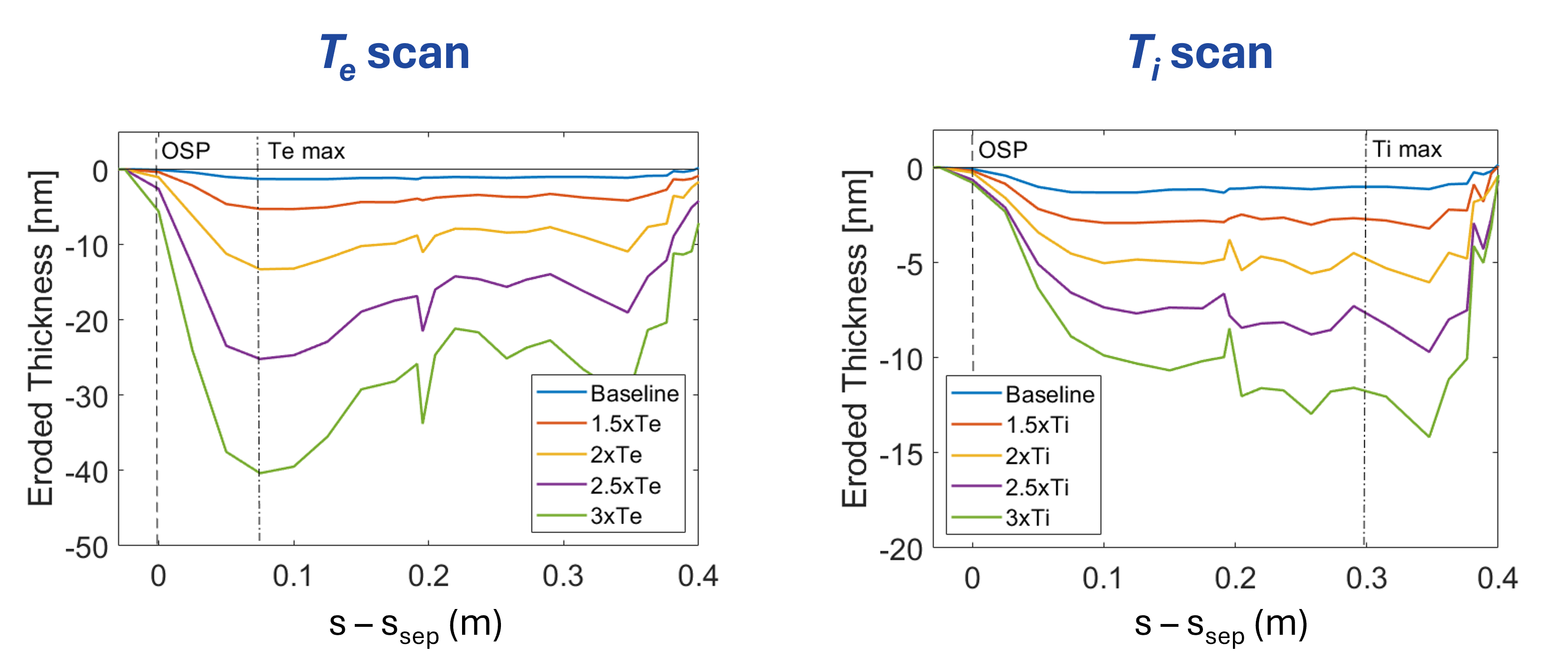}
\caption{Modelled net erosion profiles along AUG outer divertor in the proximity of the strike-point (OSP) after a fluence of $10^{24}$ He m$^{-2}$, as function of the distance from OSP location. The effect of electron temperature $T_e$ (left) and ion temperature $T_i$ (right) variations are shown. Dashed lines denoted with $T_{e,max}$ and $T_{i,max}$ represent the position with respect to OSP where the corresponding temperature peaks.}
\label{fig:AUG_Tscan}
\end{figure}

\begin{figure}
\centering
\includegraphics[width=1.0\textwidth]{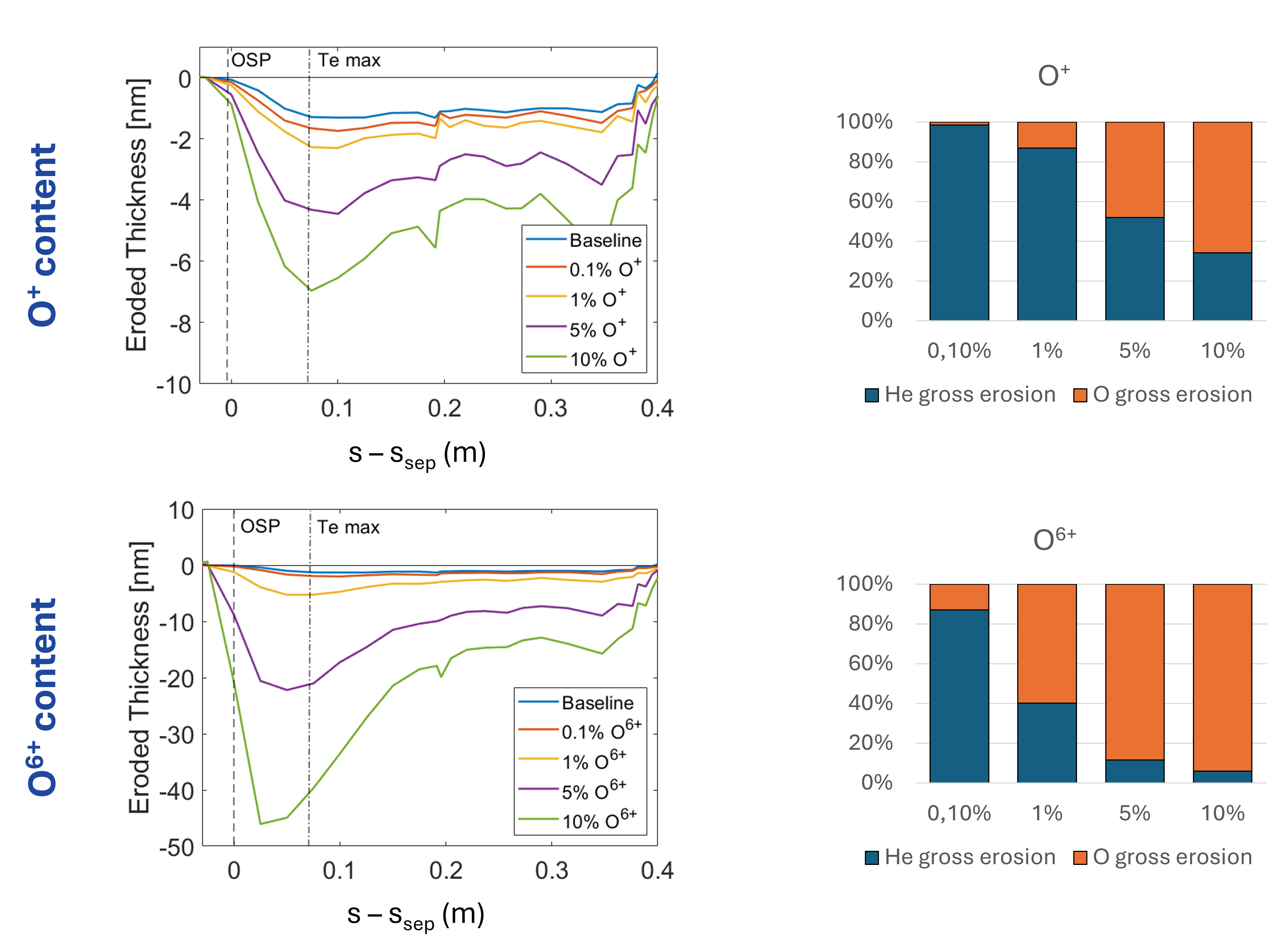}
\caption{Effect of oxygen inclusion on net erosion profiles along AUG outer divertor in the proximity of the strike-point (OSP) after a fluence of $10^{24}$ He m$^{-2}$, as function of the distance from OSP location. Both low (O$^+$, first row) and high charge state (O$^{6+}$, second row) are shown, together with O contribution to the overall outer divertor erosion in the two cases. }
\label{fig:AUG_Oscan}
\end{figure}

\begin{figure}
\centering
\includegraphics[width=0.7\textwidth]{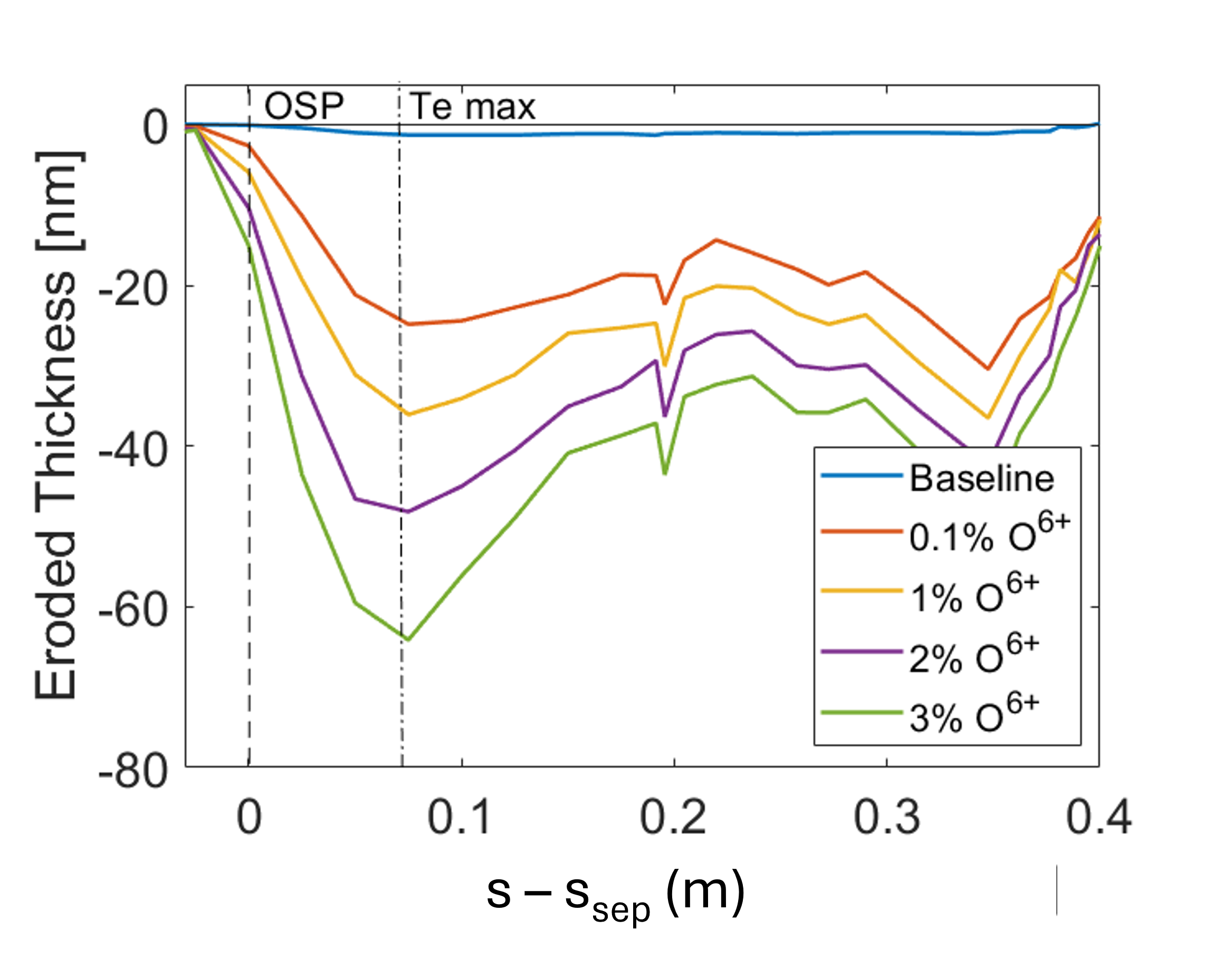}
\caption{Modelled outer divertor net erosion profile as function of O content in plasma in its high charge state (O$^{6+}$). A factor $2$ to both plasma temperatures is assumed.}
\label{fig:AUG_ExpBenchmark}
\end{figure}

Investigating fuzz formation and erosion in the tokamak environment was among the main objectives of 2019-2022 He plasma campaigns in AUG \cite{hakola2024helium, rasinski2023fib}. To this end, several W samples either pristine or with pre-formed fuzz structures on surface have been exposed to a sequence of L- and H-mode plasma discharges during a single session, exploiting AUG outer divertor manipulator \cite{vuoriheimo2024divertor,rasinski2023fib}. Focused ion beam (FIB) line marking has been used to measure sample erosion or fuzz growth after the campaign \cite{rasinski2023fib}. Ideally, one should have had two different experiments on separate days, one in L-mode and the other in H-mode. However, due to limited time, both sub-experiments were combined in a single shot day. To avoid harmful effect by the subsequent exposures, the L- and H-mode strike points were separated as much as possible, i.e. by approximately 6-7 cm, with the L-mode OSP located in the private flux region of H-mode scenario. In this way, the erosion of samples located close to the L-mode OSP, assessed at the end of the campaign by post-mortem measurements, can be mostly ascribed to this scenario \cite{hakola2024helium,rasinski2023fib,vuoriheimo2024divertor}. The region between L- and H-mode strike-points is, instead, characterised by a combination of L-mode erosion and H-mode deposition \cite{vuoriheimo2024divertor}, which complicates drawing conclusions in that location. The measured net erosion in the region close to L-mode strike-point after a fluence of $10^{24}$ He m$^{-2}$ is of the order of a few tens of nm, in the range $50-100$ nm \cite{rasinski2023fib,vuoriheimo2024divertor}. This value refers to pristine W samples, since evaluating the eroded amount from pre-damaged ones with fuzz was too difficult due to its inhomogeneity \cite{rasinski2023fib}. Reproducing this eroded thickness in ERO2.0 simulations would provide a first indication of the accuracy of this modelling activity and could also support the interpretation of experiments. \\
A first parameter which can influence estimating erosion in ERO2.0 is the incidence angle assumed for plasma ions. As previously mentioned, a fixed 80° incidence angle has been adopted in the presented ERO2.0 simulations, to account for particle deflection in the sheath from the initial magnetic field inclination (87°-88°). However, this parameter is usually not completely known from experiments and a distribution of incidence angles is expected. It is thus important to investigate how much the assumed value of the incidence angle impacts on erosion results. \\
A parametric scan varying only the incidence angle in the range 72°-84° is performed with ERO2.0, by taking the helium ion fraction from SOLPS-ITER. Outer divertor gross erosion as function of plasma ion incidence angle is presented in figure \ref{fig:AUG_HePlasmaResults}(b). One can see that incidence angle has a notable impact on gross erosion estimation, especially in this close to the grazing incidence range. In addition, in the real case, morphology could significantly influence the angular distribution of impinging species, adding further uncertainty to the picture \cite{eksaeva2021impact}. A possible way to shed light on these aspects is trying to compare these results with experiments. \\
Net erosion after the experimental fluence of $10^{24}$ He m$^{-2}$ has been evaluated in ERO2.0 simulations as a function of the incidence angle of the impinging He particles. However, even in the case featuring the highest predicted erosion, i.e. for 72° incidence, net erosion in the proximity of the strike-point is below 1 nm, still considerably lower than experimental. Moreover, extending incidence angle scan in range 0°-60°, predicted erosion is at least a factor $\sim 40$ lower with respect to experiments, represented as the baseline case in figure \ref{fig:AUG_Tscan}. \\
To understand this discrepancy, it is needed to improve present modelling considering possible experimental and model unknowns and uncertainties. SOLPS-ITER results, indeed, just provide a steady-state in time value for each plasma quantity and do not account for transients. LP measurements are affected by noise and oscillation, and SOLPS-ITER simulations were matched to their median values (see figure \ref{fig:SOLPSexp}.b). Moreover, ion temperature predicted by SOLPS-ITER is not directly validated against experimental data, i.e. it is the outcome of the simulation validated as discussed in section \ref{sec:SOLPSvalidation}. Significant uncertainties could thus be expected. Last but not least, only pure helium plasma has been considered so far, while the presence of other common impurities in plasma, such as oxygen, could play a role, as already observed in previous AUG discharges \cite{hakola2017ero}. The remaining part of this section will be thus devoted to investigate the effect of these parameters on divertor erosion. It is worth mentioning that another important uncertainty source could be represented by SDTrimSP sputtering yields, especially in the case of redeposited W, which could feature quite different adhesive properties with respect to pristine one. This is particularly relevant for the divertor region, where the overall net erosion is highly dependent on the strong prompt redeposition ($90-95\%$ in our ERO2.0 simulations). The assessment of sputtering yield variation due to redeposited W is left to future studies. \\
To address the impact of plasma temperatures variation, a multiplying factor in range $1.5-3.0$ has been applied individually to both $T_e$ and $T_i$, keeping all other simulation parameters fixed. The net erosion profiles along outer divertor in the proximity of the strike-point after the identified experimental fluence, as function of the distance from OSP location, are presented in figure \ref{fig:AUG_Tscan}. The notch around $0.2$ m refers to a discontinuity in the 3D geometry employed in this work and should be considered as an artifact. One can see that the experimental eroded thickness could only be approached considering the highest factor for $T_e$, while it could not be reached for $T_i$. This can be understood considering how the two temperatures are included in the ion impact energy formula. $T_e$ is multiplied by a factor $3q$, which is $3$ times higher than $T_i$ factor for He$^{2+}$ ions. Moreover, $T_e$ contributes more in correspondence of the OSP, that is where it reaches the highest values, as depicted in figure \ref{fig:EROinput}(b), while $T_i$ has a greater impact far from it. $T_e$ variation within its experimental uncertainty is thus a more probable candidate to partially explain the discrepancy between ERO2.0 predictions and experimental data rather than $T_i$. However, the multiplication factor needed to find a good agreement with experimental erosion is too high to be considered realistic. 
 \\
A second candidate could be the presence of impurities in plasma. To study this effect, oxygen (O) has been considered as proxy for light impurities, as it is a typical element that could be present in tokamak plasmas. Its concentration has been included in the background plasma as a percentage of the electron density $n_e$, leading to a higher O content where $n_e$ peaks, namely in the proximity of the OSP \cite{di2021modelling}. In reality, local variations could be expected, which could alter erosion results. However, their inclusion would further complicate the interpretation of present results, and it is left to future studies. As previously mentioned, also the presence of high-Z impurities, such as W coming from the main chamber is neglected here. A parametric scan over a wide range of O concentrations has been performed, from $0.1$ to $10\%$. Typical concentrations in tokamak, including AUG, are of the order of few percent, thus the $10\%$ value can be considered as an upper limit \cite{kallenbach2009non,di2021modelling,dux2011main}. Again, plasma parameters are assumed not to be influenced by impurity presence. \\
Oxygen can be ionised in different charge states in plasma, up to O$^{8+}$ when fully ionised, which can differently impact erosion results. To account for this, the parametric scan is repeated for a low charge state (O$^+$) and a high one (O$^{6+}$). The choice of the high ionization state is consistent with the experience in other tokamaks as WEST, where a good agreement between modelling and experiments could be found only considering the presence of high charge states  \cite{klepper2022}. Moreover, considering coronal equilibrium at the electron temperature of this experiment, a charge state between $5+$ and $6+$ is expected for O close to the strike-point region \cite{dufresne2021influence}. Figure \ref{fig:AUG_Oscan} shows the impact of O concentration in the two charge states on the net erosion profiles along outer divertor. The O contribution to the overall divertor erosion is also displayed on the right-hand side. Focusing on the first row, one can see that O$^+$ effect is only mild, roughly doubling the original eroded thickness at $1\%$ concentration. Even at the extreme $10\%$ concentration, the expected erosion is still approximately one order of magnitude lower than the experimental one. O$^+$ contribution is indeed still below $20\%$ for a $1\%$ concentration, equalling He erosion only with a $5\%$ content.
 \\
The situation is quite different considering O$^{6+}$ in the second row. Its impact on erosion is indeed much stronger due to the higher energy acquired by O ions in the sheath, being able to achieve an eroded thickness comparable with experimental values, even though only at the extreme $10\%$ concentration. As for $T_e$, the effect is more evident in the proximity of the OSP, that is where O concentrates with these modelling assumptions. Moreover, this is also the region where $T_e$ is higher, leading to stronger sheath acceleration. Looking at erosion contributions on the right-hand side, O is already dominating erosion in the $1\%$ case, which can be considered as a reasonable content for light impurities. \\
Up to now, erosion measurements could only be reproduced with ERO2.0 assuming extreme single plasma parameter variations, which could be hardly considered as physical. One could wonder whether it could be possible to obtain comparable results with a reasonable combination of plasma parameters. Several possibilities can be chosen to match experimental data. However, since the O content in plasma is the most uncertain parameter in these discharges, it has been decided to fix reasonable $T_e$ and $T_i$ factors with respect to LP data in figure \ref{fig:SOLPSexp}.b and consider O as a free parameter. A factor $2$ to both temperatures has thus been assumed, while O concentration in its high charge state has been varied in the range $0.1-3.0\%$. \\
The results of this scan are reported in figure \ref{fig:AUG_ExpBenchmark}. Maximum erosion is again placed in the position where $T_e$ and $n_e$ peak due to the effect of O. A reasonable agreement with experimental data can now already be found assuming an O content of a few percent, which could give an indication about the presence of light impurities in AUG discharges.

\section{Modelling W erosion and transport: fluid and kinetic approaches}
\label{sec:Wtransport}
The last part of this work is related to the comparison of ERO2.0 and SOLPS-ITER W transport predictions. Comparing different models, such as, in this case, a kinetic Monte-Carlo and a multi-fluid one, is useful to understand limitations and physical aspects which could be important in the validation process against experimental data. Moreover, including W in SOLPS-ITER could provide indications about its effect on plasma parameters, which is not accounted for in ERO2.0 simulations. ERO2.0, in turn, can be used to provide reliable estimations of W sources for SOLPS-ITER modelling. \\
\subsection{Multi-fluid simulations including W impurities: the impact of target boundary conditions}
\label{subsec:SOLPSW}
The multi-ion model implemented in SOLPS-ITER allows to consistently incorporate impurity transport within the fluid approximation. In this section, we describe the results of SOLPS-ITER simulations by adding W impurities to the He main plasma. The general simulation setup is the same as described in~\ref{app:SOLPSsetup} and in Section~\ref{sec:SOLPSvalidation} for pure-He simulations. W is included using the ‘jett’ bundling model tested in~\cite{Bonnin2011} and applied to EAST modelling in~\cite{Gao2021}, thus leading to a total of 27 plasma species, i.e. 3 helium (He) species and 24 bundled tungsten species. This bundling scheme has been chosen since it better reproduces, among all possible models tested in \cite{Bonnin2011}, the SOLPS-ITER simulation with all W charge states. For each ion species, the density and momentum equations are solved independently, while a single ion temperature is considered. Physical sputtering yields for both He-W sputtering and W-W self-sputtering are computed according to~\cite{item_2131245}. A minimum residual density equal to $1.0 \times 10^8$ m$^{-3}$ has been assumed for each bundled species for numerical stability of the fluid equations. \\
An important result of this analysis is related to the impact that the choice of boundary conditions for W velocity in front of the divertor targets has on the sputtering evaluation. Two choices are available in SOLPS-ITER. The standard hypothesis, and the only available option when drift velocities are included in the model, is to assume perfect impurity entrainment in the main ion flow. This means setting the same velocity at the magnetic pre-sheath entrance for both main ions and impurities, namely the \textit{collective fluid velocity} $v_{\parallel , coll}$ given by 
\begin{equation}
v_{\parallel , coll} = \sqrt{\frac{\sum_a [n_a (Z_a T_e + T_i)]}{\sum_a (m_a n_a)}} . 
\label{eq:v_coll}
\end{equation} 
Its value is lower but similar to the main and lighter ion sound speed and, as a result, this velocity is much higher than the W sound speed. \\
The second option consists of imposing at the magnetic pre-sheath entrance a velocity equal to the ion sound speed for each species, namely the \textit{single species fluid velocity} $v_{\parallel , a}$ given by
\begin{equation}
v_{\parallel , a} = \sqrt{\frac{(Z_a T_e + T_i)}{m_a}} . 
\label{eq:v_W}
\end{equation}
In this case, $v_{\parallel , He} >> v_{\parallel , W}$, due to the much higher mass of W ions with respect to He ions and their consequently higher inertia. \\
Gross tungsten source from the inner and outer divertor targets resulting from SOLPS-ITER simulations using boundary conditions~\ref{eq:v_W} and~\ref{eq:v_coll} are reported in the first two lines of table~\ref{tab:erosion}, respectively. These results highlight the impact that fluid boundary conditions have on the evaluation of the W source. Indeed, using the collective velocity boundary condition produces a $\sim 20$ times higher W source than the species dependent sound speed boundary condition. This striking difference is almost entirely due to the greatly amplified effect of W self-sputtering that is obtained when W ions are accelerated to almost helium sound speed at the magnetic pre-sheath entrance.  \\
It is worth pointing out that SOLPS-ITER inner divertor profiles are not benchmarked against experimental data, which would require the use of drifts in SOLPS. However, since the main aim here is a comparison between codes, a proper benchmark is out of the scope of this section.

\begin{table}[h!]
\centering
\arrayrulecolor{gray}
\begin{tabular}{|l|c|c|}
\hline
 [W atoms/s] & $\Gamma_{W spt, TOT}$ & $\Gamma_{W spt, SELF}$ \\
\hline 
SOLPS-ITER ($v_{\parallel, W}$) & $5.08 \times 10^{19}$ & $2.03 \times 10^{19}$ \\
\hline
SOLPS-ITER ($v_{\parallel, Coll}$) & $90.08 \times 10^{19}$ & $83.96 \times 10^{19}$ \\
\hline
ERO2.0 & $3.51 \times 10^{19}$ & $0.34 \times 10^{19}$ \\
\hline
Analytical (80°) & $0.10 \times 10^{19}$ & - \\
\hline
Analytical (60°) & $1.05 \times 10^{19}$ & - \\
\hline
Analytical (0°) & $1.18 \times 10^{19}$ & - \\
\hline
\end{tabular}
\caption{Total ($\Gamma_{W spt, TOT}$) and self W sputtering ($\Gamma_{W spt, SELF}$) from the outer and inner divertor targets computed with different models. All reported data refer to gross erosion rates, to eliminate any variation due to W transport between SOLPS-ITER and ERO2.0.}
\label{tab:erosion}
\end{table}

\subsection{Evaluating the tungsten source: fluid, kinetic and analytical results compared} \label{subsec:erosion_comparison}
 It is now essential to identify which of the two boundary conditions discussed in section~\ref{subsec:SOLPSW} produces comparable results with respect to ERO2.0. To this purpose, the one computed by ERO2.0 can be used as the reference W source since it does not need any boundary condition at the sheath entrance and the velocity that the different species have in front of the targets results directly from following the individual kinetic trajectories. SOLPS-ITER evaluates W sputtering from both inner and outer targets, while previous ERO2.0 modelling only focused on the outer divertor. Thus, dedicated ERO2.0 simulations have been performed with the same configuration of W sources as in the SOLPS-ITER one. The He incidence angle in ERO2.0 is obtained by sampling the 3D velocity vector at the sheath entrance, calculated from SOLPS-ITER plasma parameters, by Maxwellian distributions, which is expected to be the most similar assumption with respect to the EIRENE one. No temperature variation or foreign impurity content has been assumed for this simulation. The overall divertor W source found in ERO2.0 is reported in table~\ref{tab:erosion}, in comparison to the SOLPS-ITER ones. All reported data refer to gross erosion rates, to eliminate any variation due to W transport. ERO2.0 predicts a W source that is clearly more similar to the one estimated by SOLPS-ITER in the species dependent Bohm velocity boundary condition~\ref{eq:v_W}.  Compared to ERO2.0, thus, the perfect entrainment assumption, routinely used in SOLPS-ITER simulations with impurities, which imposes a collective velocity given by equation~\ref{eq:v_coll} at the magnetic pre-sheath entrance, leads to a significant overestimation of the self-sputtering contribution. 
\begin{figure}
\centering
\includegraphics[width=0.8\textwidth]{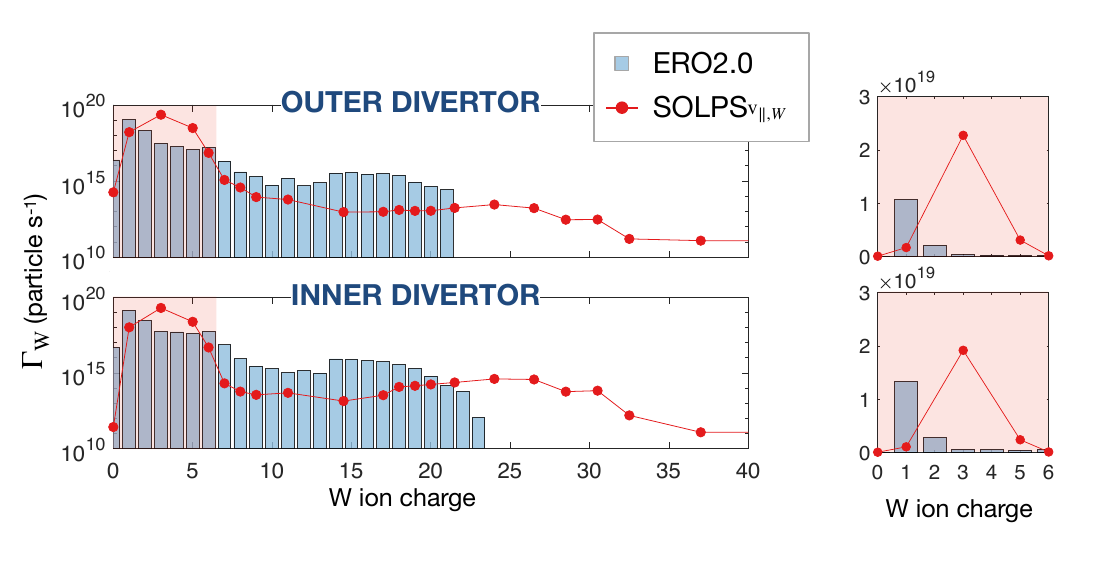}
\caption{Charge distribution of W particles impinging on the outer and inner divertor targets in ERO2.0 (barplot) and SOLPS-ITER simulations (red dots)  with species dependent Bohm velocity boundary condition~\ref{eq:v_W}. The red boxes on the right highlight the impact of prompt redeposition, included in ERO2.0 but not in SOLPS-ITER.}
\label{fig:AUG_ERO-SOLPSerosion}
\end{figure} \\
By comparing the ERO2.0 and SOLPS-ITER solutions under the species dependent sound speed condition, almost the entire difference can be ascribed again to a different self-sputtering source, while He sputtering contribution is almost identical. To investigate this discrepancy, figure \ref{fig:AUG_ERO-SOLPSerosion} reports the charge distribution of W impinging onto the two targets in the two codes. We can clearly notice that the ion distribution in the SOLPS-ITER case is shifted towards higher charges compared to ERO2.0. Looking at the boxes on the right, we see that the SOLPS-ITER distribution is peaked on the W-bundle $\{+2, +3, +4\}$, while the one from ERO2.0 is peaked on the W$^+$ charge state. The underestimation of W$^+$ flux in SOLPS-ITER results in an overestimation of the second lower charge state, i.e. the W-bundle $\{+2, +3, +4\}$. This effect can be attributed to the absence of a prompt redeposition model in SOLPS, which is instead consistently evaluated in ERO2.0. Ultimately, because of the absence of prompt redeposition, SOLPS-ITER finds a W flux impinging on the targets with a higher impact energy on average. This is the effect of a stronger electrostatic acceleration in the sheath associated with the - on average - higher charge states, thus leading to an overestimation of the self-sputtering contribution. In addition, self-sputtering is enhanced by the stronger high charge state tail of the SOLPS-ITER distribution with respect to ERO2.0 one. This could be partially explained, on the one hand, by the residual density imposed in SOLPS-ITER for each bundled W species, which leads to a minimum flux of about $10^{12}$ particles s$^{-1}$ even for high charge states. On the other one, the very low fluxes of high charge states could be hardly resolved by the Monte-Carlo method employed in ERO2.0 due to statistical reasons (nearly zero test particles with high charge out of several millions). \\
As a final comparison, W sources have been evaluated exploiting the analytical model described in section~\ref{sec:analyticalModel}. Self-sputtering cannot be estimated with this model, thus its total sputtered W flux should be compared to $\Gamma_{W spt, TOT} - \Gamma_{W spt, SELF}$ of both SOLPS-ITER and ERO2.0. The saturation current $j_{sat}$ has been retrieved from the validated SOLPS-ITER simulation, integrating it on both targets. An electron temperature of 22 eV has been assumed, comparable to the one reported on the basis of SOLPS-ITER modelling in figure \ref{fig:SOLPSexp}. The He ion incidence angle is varied from 0° (orthogonal incidence) to 80°, with 60° used as a standard assumption to account for sheath deflection. The $\Gamma_{W spt, TOT}$ obtained with these assumptions is reported in table \ref{tab:erosion}. One can see a consistently lower erosion in the 80° case with respect to less shallow angles, which are, instead, in good agreement with both SOLPS-ITER and ERO2.0 results. This is due to the rather strong assumption of fixed incidence angle in the analytical model, compared to the distributions employed in the two codes in these simulations. Thus, for a proper analytical estimation of He erosion in the divertor region with fixed incidence angle, one would better consider strong deflections with respect to the magnetic field incidence rather than very shallow angles.

\subsection{Impact of W anomalous diffusivity: SOLPS-ITER vs. ERO2.0 treatment}
\label{subsec:SOLPS-ERO_Wdensity}
\begin{figure}
\centering
\includegraphics[width=0.8\textwidth]{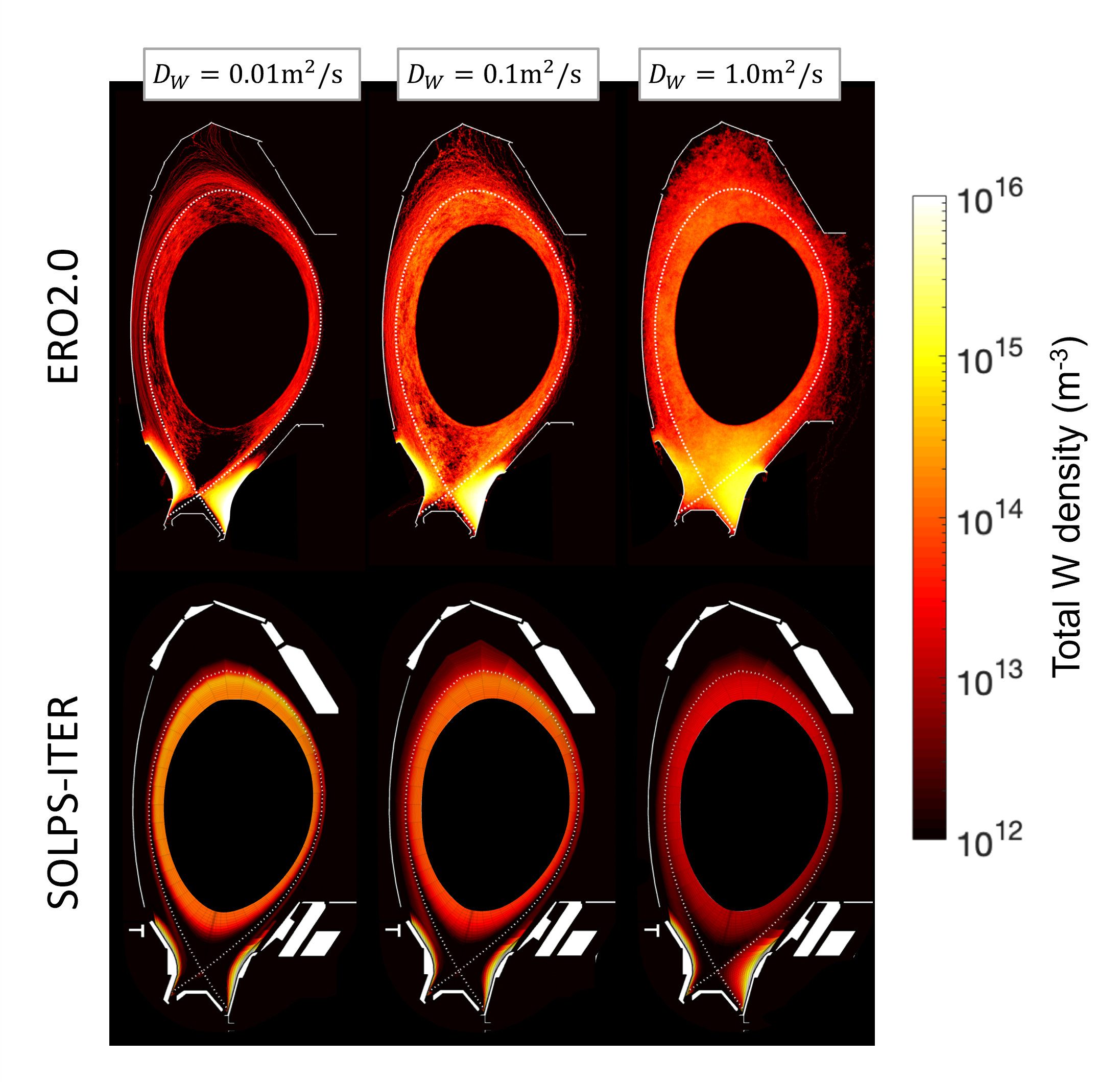}
\caption{Poloidal distribution of W density as a function of the anomalous diffusion coefficient $D_W$, from ERO2.0 (first row) and SOLPS-ITER (second row) simulations. }
\label{fig:AUG_ERO-SOLPSdensity}
\end{figure}

Finally, once verified which SOLPS-ITER assumption allows to reasonably reproduce ERO2.0 W source, it is now possible to compare the two W transport models in terms of their dependence on anomalous diffusivity. The aim of this comparison is running the two codes independently but employing a comparable W source. \\
Both codes employ diffusion coefficients $D_W$ to account for anomalous cross-field transport, but with a different model. SOLPS-ITER assumes a purely diffusive radial transport with no drifts, which can be written in the form of Fick's law:
\begin{equation}
\Gamma_{r,W} = - D_W \frac{\partial n_W}{\partial r} .
\end{equation}
In ERO2.0 drifts are self-consistently included in the kinetic description and the code accounts for anomalous diffusion by adding a radial contribution at the end of each time step, proportional to the square root of time since previous collision with plasma or wall $\tau_{coll}$, i.e.
\begin{equation}
\Delta r_W \propto \sqrt{D_W \tau_{coll}} .
\end{equation} 
 To understand the effect of this parameter in the two cases, different simulations have been performed considering a variable $D_W$ in the range $0.01-1.0$ m$^2$ s$^{-1}$. In all cases, the single species boundary condition for the sound speed is employed in SOLPS-ITER, while no multiplication factor nor oxygen content is assumed in ERO2.0, in accordance with the W sources reported in table \ref{tab:erosion}. The poloidal distribution of W density in these simulations is reported in figure \ref{fig:AUG_ERO-SOLPSdensity}. Focusing on the two cases with $D_W = 1.0$ m$^2$ s$^{-1}$, one can see important differences, both in the proximity of the X-point and in the core. Transport towards the X-point appears to be stronger in ERO2.0, leading to a higher W density near the X-point and, accordingly, in the core region. This can be explained by considering the impact of the $\nabla B$ drift, whose effect is included in ERO but not in the present SOLPS-ITER simulation. Indeed, the $\nabla B$ drift, in reversed magnetic field discharges as the one considered in this work, is directed towards the X-point. More similar W distribution can be found by reducing $D_W$ in ERO2.0 by a factor 100, employing $D_W = 0.01$ m$^2$ s$^{-1}$ (see top-left and bottom-right boxes in figure~\ref{fig:AUG_ERO-SOLPSdensity}). A lower $D_W$ in ERO2.0, indeed, reduces W penetration through the separatrix, which occurs primarily near the X-point. \\
It is now interesting to investigate what is the effect of decreasing $D_W$ in SOLPS-ITER. As in ERO2.0, a lower $D_W$ leads to a lower W diffusion from the targets towards the X-point. In core, instead, the situation is almost opposite, with a W density increase according to SOLPS. The observed W core accumulation is influenced by core boundary conditions (BCs) for W, which act as a W source there in the SOLPS-ITER case. Indeed, the poloidally averaged density at the core boundary ring for each bundled species is obtained by setting the corresponding particle flux to zero. A better BC, for future works, could be to prescribe to 0 the total W ion flux summed over all bundles, allowing for non-zero fluxes of the single bundles. To do that, however, one would need to know the maximum W charge state exiting the core ($W^\mathrm{max}$), so that the total inward W flux could be recycled at the SOLPS-ITER core boundary as $W^\mathrm{max}$. In the absence of this piece of information, imposing null flux for each W bundled species across the core boundary is necessary, leading to artificial W source at the core boundary. When $D_W$ is low, outward radial transport is also low and an overall higher W density inside the separatrix is observed. \\
A further possibility for SOLPS-ITER BCs could be to impose the density of each bundle by using a Dirichlet condition that sets the W density retrieved from ERO2.0 simulations. Indeed, as observed in section~\ref{subsec:SOLPSW}, in ERO2.0 particles are traced throughout their trajectory without solving conservation equations. All particles reaching the core boundary are teleported in a random location close to it (1 cm in these simulations) in both poloidal and toroidal direction, without introducing any source. The investigation of the impact of this and other core BCs on W migration in SOLPS-ITER could be of interest for future studies.
 \\
Once fixed a similar W source as in this work, the different behaviour of the two models with respect to anomalous diffusivity is overall a result of their peculiar assumptions. Both codes, indeed, rely on approximations which could partially miss the full physical picture.  SOLPS-ITER assumes fluid population for each bundled W species, which presumes a sufficient impurity density to act as a fluid. This is almost the opposite with respect to ERO2.0 test particle approximation, which assumes W particles being sufficiently sparse to influence neither each other nor the background plasma. In general, as soon as W accumulates, e.g. due to low cross-field transport, ERO2.0 could be missing spreading effects due to W-W interaction, especially at a local scale. On the contrary, whenever W density remains well below the main plasma ion density, the SOLPS-ITER fluid approach may become unrealistic with respect to the full kinetic ERO2.0 transport. Including drifts in SOLPS-ITER simulations may reduce the discrepancies observed between the two codes, especially close to the X-point. Other mechanisms which could play a role on W density calculations are thermal forces and neoclassical pinch velocity, which could be implemented in ERO2.0 \cite{di2023first,kumpulainen2024validated}. However, these are not included in present simulations and their impact on W migration could be investigated in future studies.  

\section{Conclusions}
\label{sec:conclusions}
This work presented a thorough modelling effort to simulate plasma-wall interaction in L-mode helium plasma discharges in AUG, by means of a simple analytical model and the SOLPS-ITER and ERO2.0 codes. The main aims were estimating the impact of He ionization state on W divertor erosion and investigating the differences between fluid and kinetic approaches in the modelling of W sources and transport in He plasmas. \\
For the first objective, a simple analytical model has been developed, showing an opposite behaviour of W erosion with increased fraction of He$^{2+}$ in the plasma depending on the plasma temperature. These findings have then been successfully tested in the realistic environment of AUG He plasma discharges, using SOLPS-ITER solutions as the background plasma. The case with the estimated SOLPS-ITER He ion fraction in plasma has been compared to the two extreme assumptions of pure He$^+$ and pure He$^{2+}$ plasma, showing that the latter features a slightly higher divertor erosion with respect to the most accurate case, while the former yields a lower erosion by orders of magnitude. \\
Obtained results in He plasma, in terms of eroded thickness, have been compared to post-mortem experimental data. Even considering the ERO2.0 parameters causing the highest erosion rates with the provided inputs, the code is shown to underestimate divertor erosion by more than one order of magnitude. Thus, possible unknowns about experimental conditions have been analysed, such as the impurity level in the plasma (using oxygen as proxy for light impurities) and the uncertainty about temperature measurements. A variation of one single parameter, namely electron temperature, ion temperature or oxygen content, is not sufficient to satisfactorily explain alone the discrepancy with post-mortem data. These observations are not only relevant for He plasma discharges, but could also be considered as general for the tokamak environment. Thus, improving the accuracy in measuring those parameters, with particular reference to the impurity level in plasma, could be essential for evaluating the PFC erosion. In addition, estimating the impact of redeposited W or surfaces with strong modifications or increased roughness on SDTrimSP sputtering yield evaluation could be important to reduce the uncertainty of these simulations. \\
The inclusion of W in SOLPS-ITER modelling and the comparison with ERO2.0 results revealed the crucial role of target boundary conditions on W divertor erosion. \textit{Collective fluid velocity} condition at the magnetic pre-sheath entrance has been shown to significantly overestimate W self-sputtering contribution. The absence of prompt redeposition in SOLPS-ITER has been identified as the main responsible of the smaller discrepancies between ERO2.0 and SOLPS-ITER W sources with \textit{single species fluid velocity} condition, leading to a higher self-sputtering contribution in the latter. \\ 
Finally, SOLPS-ITER and ERO2.0 W transport predictions in He plasma have been compared, describing the peculiarities of the two models in terms of their dependence on anomalous diffusivity variation. On this side, the comparison with experimental data of W density in the volume, e.g. bolometry measurements, would probably allow a better understanding of the reliability and applicability domain of the two models. \\
The natural continuation of this activity in AUG He plasma would be the modelling of H-mode discharges, which would be also associated with a higher amount of experimental data provided by markers located near the H-mode strike-point. Theoretical modelling of H-mode discharges with available tools is currently ongoing to complement present results.

\section*{Acknowledgements}
G. Alberti, M. Passoni and C. Tuccari acknowledge funding from Eni SpA. \\
This work has been carried out within the framework of the EUROfusion Consortium, partially funded by the European Union via the Euratom Research and Training Programme (Grant Agreement No 101052200 — EUROfusion). The Swiss contribution to this work has been funded by the Swiss State Secretariat for Education, Research and Innovation (SERI). Views and opinions expressed are however those of the author(s) only and do not  necessarily reflect those of the European Union, the European Commission or SERI. Neither the European Union nor the European Commission nor SERI can be held responsible for them. \\

\appendix
\section{SOLPS-ITER simulation setup}
\label{app:SOLPSsetup}
\begin{figure}
\centering
\includegraphics[width=0.3\textwidth]{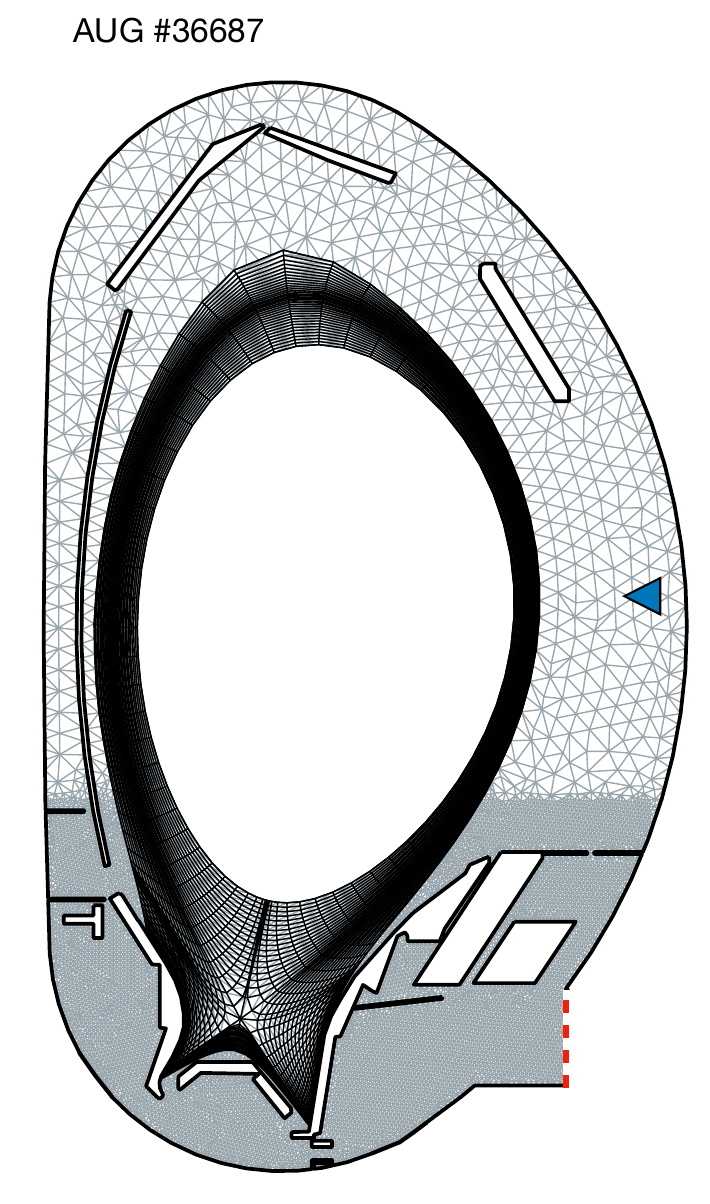}
\caption{SOLPS-ITER simulation geometry. B2.5 plasma mesh is in black and EIRENE neutral mesh in gray. Pumping surface is indicated with a red dashed line and the puffing location with the blue triangle. }
\label{fig:solpsGeo}
\end{figure}

The SOLPS-ITER simulations presented in this work were performed using v3.0.9. The geometry of these simulations is reported in figure~\ref{fig:solpsGeo}.
The plasma density, momentum and energy conservation equations are solved using the following boundary conditions:
\begin{enumerate}
\item Divertor targets: sheath boundary conditions, for densities, temperatures, velocities and electrostatic potential. The impact of the velocity boundary condition when W impurities are included in the simulation is addressed in detail in section~\ref{subsec:SOLPSW}.
\item Far SOL and private flux region (PFR): zero gradients of the parallel velocities, zero current for the potential equation and decay boundary condition for $n_a$, $T_e$ and $T_i$ with $\lambda_\mathrm{decay} = 1 \, \mathrm{cm}.$ 
\item Core: zero parallel velocity gradient and zero current boundary conditions. The total power entering the edge region from the core is estimated from experiments as $P_\mathrm{e+i, core} = P_\mathrm{Ohm} + P_\mathrm{ECRH} - P_\mathrm{rad} \simeq 1.13$ MW and it is set as boundary condition for the electron and ion energy equation assuming equal power sharing between the two populations. The $\mathrm{He^{2+}}$ particle flux is assumed equal and opposite to the neutral He flux reaching the core boundary ($\Gamma_\mathrm{core, B2.5, He^{2+}} + \Gamma_\mathrm{core, EIRENE, He} = 0$) while zero flux is imposed for $\mathrm{He^{+}}$.

\end{enumerate}

At the walls, all particles are fully recycled as fast (according to the TRIM database for He projectiles on W recycling surface) or thermal (at wall temperature) He atoms. At the location corresponding to the turbo-molecular pump (dashed red line in figure~\ref{fig:solpsGeo}) an absorption probability $p_a = 0.007$ is set for He atoms \cite{zito2021numerical}. Helium gas at ambient temperature is injected from the OMP, as indicated by the blue triangle in figure~\ref{fig:solpsGeo}. \\
Electron density in the simulation is controlled by a feedback scheme that adjusts the He puffing strength to get to the desired separatrix electron density at the OMP $n_{e,\mathrm{sep@OMP}} = 1.25 \times 10^{19} \, \mathrm{m^{-3}}$. The atomic processes included in EIRENE are the ones listed in Table 1 in~\cite{Alberti2023}.

\bibliographystyle{iopart-num}
\bibliography{Mybib} 

\providecommand{\newblock}{}
\begin{thebibliography}{10}
\expandafter\ifx\csname url\endcsname\relax
  \def\url#1{{\tt #1}}\fi
\expandafter\ifx\csname urlprefix\endcsname\relax\def\urlprefix{URL }\fi
\providecommand{\eprint}[2][]{\url{#2}}

\bibitem{roth2009}
Roth J, Tsitrone E, Loarte A, Loarer T, Counsell G, Neu R, Philipps V, Brezinsek S, Lehnen M, Coad P {\em et~al.\/} 2009 {\em Journal of nuclear materials\/} {\bf 390} 1--9

\bibitem{smirnov2015}
Smirnov R, Krasheninnikov S, Pigarov A~Y and Rognlien T 2015 {\em Physics of Plasmas\/} {\bf 22}

\bibitem{loarer2009}
Loarer T 2009 {\em Journal of Nuclear Materials\/} {\bf 390} 20--28

\bibitem{hakola2017}
Hakola A, Brezinsek S, Douai D, Balden M, Bobkov V, Carralero D, Greuner H, Elgeti S, Kallenbach A, Krieger K {\em et~al.\/} 2017 {\em Nuclear fusion\/} {\bf 57} 066015

\bibitem{baldwin2010}
Baldwin M and Doerner R 2010 {\em Journal of Nuclear Materials\/} {\bf 404} 165--173

\bibitem{tsitrone2022}
Tsitrone E, Pegourie B, Gunn J, Bernard E, Bruno V, Corre Y, Delpech L, Diez M, Douai D, Ekedahl A {\em et~al.\/} 2022 {\em Nuclear Fusion\/} {\bf 62} 076028

\bibitem{reinhart2022}
Reinhart M, Brezinsek S, Kirschner A, Coenen J, Schwarz-Selinger T, Schmid K, Hakola A, Van Der~Meiden H, Dejarnac R, Tsitrone E {\em et~al.\/} 2022 {\em Nuclear Fusion\/} {\bf 62} 042013

\bibitem{romanelli2024}
Romanelli F 2024 {\em Nuclear Fusion\/}

\bibitem{wiesen2015}
Wiesen S, Reiter D, Kotov V, Baelmans M, Dekeyser W, Kukushkin A, Lisgo S, Pitts R, Rozhansky V, Saibene G, Veselova I and Voskoboynikov S 2015 {\em Journal of Nuclear Materials\/} {\bf 463} 480--484 ISSN 00223115 \urlprefix\url{https://linkinghub.elsevier.com/retrieve/pii/S0022311514006965}

\bibitem{bonnin2016}
BONNIN X, DEKEYSER W, PITTS R, COSTER D, VOSKOBOYNIKOV S and WIESEN S 2016 {\em Plasma and Fusion Research\/} {\bf 11} 1403102–1403102 ISSN 1880-6821 \urlprefix\url{http://dx.doi.org/10.1585/pfr.11.1403102}

\bibitem{romazanov2017}
Romazanov J, Borodin D, Kirschner A, Brezinsek S, Silburn S, Huber A, Huber V, Bufferand H, Firdaouss M, Br\"{o}mmel D, Steinbusch B, Gibbon P, Lasa A, Borodkina I, Eksaeva A and Linsmeier C 2017 {\em Physica Scripta\/} {\bf T170} 014018 ISSN 1402-4896 \urlprefix\url{http://dx.doi.org/10.1088/1402-4896/aa89ca}

\bibitem{fischer2010integrated}
Fischer R, Fuchs C, Kurzan B, Suttrop W, Wolfrum E and Team A~U 2010 {\em Fusion science and technology\/} {\bf 58} 675--684

\bibitem{TonelloPhD2023}
Tonello E 2023 {\em Modelling of boundary plasmas in linear devices and tokamaks\/} Phd thesis Politecnico di Milano available at \url{https://hdl.handle.net/10589/196383}

\bibitem{stangeby2000plasma}
Stangeby P~C {\em et~al.\/} 2000 {\em The plasma boundary of magnetic fusion devices\/} vol 224 (Institute of Physics Pub. Philadelphia, Pennsylvania)

\bibitem{behrisch2007sputtering}
Behrisch R 2007 {\em Sputtering by particle bombardment: Experiments and computer calculations from threshold to MeV energies\/} (Springer Science \& Business Media)

\bibitem{mutzke2024sdtrimsp}
Mutzke A, Toussaint U~v, Eckstein W, Dohmen R and Schmid K 2024

\bibitem{romazanov2024validation}
Romazanov J, Brezinsek S, Baumann C, Rode S, Kirschner A, Wang E, Effenberg F, Borodin D~V, Navarro-Gonzalez M, Xie H {\em et~al.\/} 2024 {\em Nuclear Fusion\/}

\bibitem{rode2022}
Rode S, Romazanov J, Reiser D, Brezinsek S, Linsmeier C and Pukhov A 2022 {\em Contributions to Plasma Physics\/} {\bf 62} e202100172

\bibitem{romazanov2021sensitivity}
Romazanov J, Brezinsek S, Pitts R, Kirschner A, Eksaeva A, Borodin D, Veshchev E, Neverov V, Kukushkin A, Alekseev A {\em et~al.\/} 2021 {\em Nuclear materials and energy\/} {\bf 26} 100904

\bibitem{rode2024multi}
Rode S, Brezinsek S, Groth M, Kirschner A, Matveev D, Moser L, Pitts R, Romazanov J, Terra A, Wauters T {\em et~al.\/} 2024 {\em Nuclear Fusion\/} {\bf 64} 086032

\bibitem{hakola2024helium}
Hakola A, Balden M, Baruzzo M, Bisson R, Brezinsek S, Dittmar T, Douai D, Dunne M, Garzotti L, Groth M {\em et~al.\/} 2024 {\em Nuclear Fusion\/} {\bf 64} 096022

\bibitem{rasinski2023fib}
Rasi{\'n}ski M, Brezinsek S, Kreter A, Dittmar T, Krieger K, Balden M, de~Marne P, Dux R, Faitsch M, Hakola A {\em et~al.\/} 2023 {\em Nuclear Materials and Energy\/} {\bf 37} 101539

\bibitem{vuoriheimo2024divertor}
Vuoriheimo T, Hakola A, Likonen J, Krieger K, Balden M, Radovi{\'c} I~B, Provatas G, Siketi{\'c} Z, Nizi{\'c} K~I, Rasinski M {\em et~al.\/} 2024 {\em Nuclear materials and energy\/} {\bf 41} 101766

\bibitem{eksaeva2021impact}
Eksaeva A, Borodin D, Romazanov J, Kirschner A, Kreter A, G{\"o}ths B, Rasinski M, Unterberg B, Brezinsek S, Linsmeier C {\em et~al.\/} 2021 {\em Nuclear Materials and Energy\/} {\bf 27} 100987

\bibitem{hakola2017ero}
Hakola A, Airila M~I, Mellet N, Groth M, Karhunen J, Kurki-Suonio T, Makkonen T, Sillanp{\"a}{\"a} H, Meisl G, Oberkofler M {\em et~al.\/} 2017 {\em Nuclear Materials and Energy\/} {\bf 12} 423--428

\bibitem{di2021modelling}
Di~Genova S, Gallo A, Fedorczak N, Yang H, Ciraolo G, Romazanov J, Marandet Y, Bufferand H, Guillemaut C, Gunn J {\em et~al.\/} 2021 {\em Nuclear Fusion\/} {\bf 61} 106019

\bibitem{kallenbach2009non}
Kallenbach A, Dux R, Mayer M, Neu R, P{\"u}tterich T, Bobkov V, Fuchs J, Eich T, Giannone L, Gruber O {\em et~al.\/} 2009 {\em Nuclear Fusion\/} {\bf 49} 045007

\bibitem{dux2011main}
Dux R, Janzer A, P{\"u}tterich T, Team A~U {\em et~al.\/} 2011 {\em Nuclear Fusion\/} {\bf 51} 053002

\bibitem{klepper2022}
Klepper C~C, Unterberg E~A, Marandet Y, Curreli D, Grosjean A, Harris J~H, Johnson C~A, Gallo A, Goniche M, Guillemaut C {\em et~al.\/} 2022 {\em Plasma Physics and Controlled Fusion\/} {\bf 64} 104008

\bibitem{dufresne2021influence}
Dufresne R, Del~Zanna G and Badnell N 2021 {\em Monthly Notices of the Royal Astronomical Society\/} {\bf 503} 1976--1986

\bibitem{Bonnin2011}
Bonnin X and Coster D 2011 {\em Journal of Nuclear Materials\/} {\bf 415} S488–S491 ISSN 0022-3115 \urlprefix\url{http://dx.doi.org/10.1016/j.jnucmat.2010.10.041}

\bibitem{Gao2021}
Gao S, Liu X, Deng G, Ming T, Li G, Zhang X, Tao Y and Gao X 2021 {\em AIP Advances\/} {\bf 11} ISSN 2158-3226 \urlprefix\url{http://dx.doi.org/10.1063/5.0037381}

\bibitem{item_2131245}
Eckstein W, Garcia-Rosales C, Roth J and Ottenberger W 1993 {Sputtering Data} Tech. Rep. IPP 9/82 Max-Planck-Institut f{\"u}r Plasmaphysik Garching

\bibitem{di2023first}
Di~Genova S, Ciraolo G, Gallo A, Romazanov J, Fedorczak N, Bufferand H, Tamain P, Rivals N, Marandet Y, Brezinsek S {\em et~al.\/} 2023 {\em Nuclear Materials and Energy\/} {\bf 34} 101340

\bibitem{kumpulainen2024validated}
Kumpulainen H, Groth M, Brezinsek S, Casson F, Corrigan G, Frassinetti L, Harting D, Romazanov J {\em et~al.\/} 2024 {\em Plasma Physics and Controlled Fusion\/} {\bf 66} 055007

\bibitem{zito2021numerical}
Zito A, Wischmeier M, Carralero D, Manz P, P{\'e}rez I~P, Passoni M, Team A~U {\em et~al.\/} 2021 {\em Plasma Physics and Controlled Fusion\/} {\bf 63} 075003

\bibitem{Alberti2023}
Alberti G, Tonello E, Carminati P, Uccello A, Bonnin X, Romazanov J, Brezinsek S and Passoni M 2023 {\em Nuclear Fusion\/} {\bf 63} 026020 ISSN 1741-4326 \urlprefix\url{http://dx.doi.org/10.1088/1741-4326/acacaf}

\end{thebibliography}

\end{document}